\documentstyle[psfig,art10,a4wide]{article}
\begin{document}
\parskip = 10pt
\newtheorem{corollary}{Corollary}

\begin{center}
{\Large \bf Riemann zeta function is a fractal}

{\large S.C. Woon}\\
Imperial College, Blackett Laboratory, Prince Consort Road,
London SW7 2BZ, England\\
Email: woon@vxcern.cern.ch / woon@lono.jpl.nasa.gov / s.woon@ic.ac.uk
\end{center}

\par {\narrower \narrower
\noindent {\bf Abstract} -- Voronin's theorem on the ``Universality'' of
Riemann zeta function is shown to imply that Riemann zeta function
is a fractal (in the sense that Mandelbrot set is a fractal) and
a concrete ``representation'' of the ``giant book of theorems'' that Paul
Halmos referred to.

\noindent {\bf Keywords}: Riemann zeta function, Voronin, fractal, Paul
Halmos, Info. theory \par}

Voronin's theorem \cite{Voronin} on the ``Universality'' of Riemann zeta
function \cite{Titchmarsh} states that

\noindent {\bf Theorem} $\; \;$ Let $0 < r < 1/4$ and let $f(s)$
be a complex function analytic and continuous for $|s| \leq r$.
If $f(s) \neq 0$, then for every $\epsilon > 0$, there exists a real
number $T=T(\epsilon,f)$ such that

\[ \max_{|s| \leq r} \; \left| f(s) - \zeta(s+(\frac{3}{4} + i\;T))
 \right| < \epsilon \]

Let's infer 3 Corollaries from Voronin's theorem.
The 1st is interesting, the 2nd is a strange and amusing consequence,
and the 3rd is ludicrous and shocking (but a consequence nevertheless).

\begin{corollary}
Riemann zeta function is a fractal.
\end{corollary}

\noindent {\bf Proof}\\
\indent Choose $f(s) = \zeta(a_{m}\: s+s_{0}), \; |s| \leq r, \;
a_m = a_0 + m \delta, \; \delta \ll r,$ where $m$ is a postive integer,
and $a_m$ and $\delta$ are real.

By Voronin's theorem, there exists a real number $T=T_m$ such that
\begin{equation}
|\zeta(a_{m}\: s+s_{0}) - \zeta(s+3/4+i\:T_{m})| < \epsilon
\end{equation}
Denote the disc $|a_{m}\: s| \leq a_m \: r$ centered at $s_0$ as $D_m$,
and the disc $|s| \leq r$ centered at $3/4+i\:T_m$ as $D'_m$.
Each $D_m$ has a radius dependent on $m$ while all $D'_m$ have
the same fixed radius $r$.

Define a map from $D_m$ to $D'_m$,
\[ \mu(m): \; \; (a_{m}\: s+s_{0}) \longmapsto (s+3/4+i\:T_{m}) \]
such that eqn $(1)$ holds.

Choose $m=1$; we have $T=T_1$. Now, choose $m=2$. For eqn $(1)$ to hold,
it is neccessary that $T=T_2 \neq T_1$, since $D_2$ has grown by $\delta$
in radius compared to $D_1$, while $D'_1$ and $D'_2$ have the same fixed
radius $r$. Similarly, $T_m \neq T_q$ for $m \neq q$.
This is easier to see if we draw out the schematic diagrams of the mapping
$\mu(m): \; D_m \mapsto D'_m$ as in {\bf Figure $1$}.
All the discs $D_m$ have different radii and are concentric at $s_0$, while
all the discs $D'_m$ have the same radii but are scattered and centered over
the line $3/4+i\:T_m$.
If $a_0 > 1$, as $m$ increases, larger discs are mapped by $\mu(m)$ into
smaller discs with radii r.
If $a_0 < 1$, initially, smaller discs are mapped by $\mu(m)$ into larger
discs with radii r, but as $m$ increases, discs $D_m$ will eventually become
larger than discs $D'_m$.

\begin{figure}[htbp]
\centerline{ \psfig{file=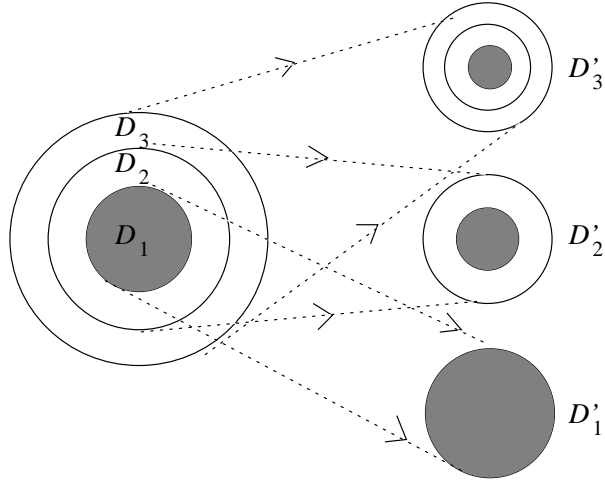}}
\caption{Map $\mu(m) : \; \; D_m \longmapsto D'_m$}
\end{figure}

Since $D_0$ is the smallest disc of all discs $D_m$, the region of $D_0$, ie.
disc $|a_0\: s| < r$ centered at $s_0$ can be found within all discs $D'_m$,
as the region of $D_0$ is mapped by $\mu(m)$ into the region $|s| < r/{a_m}$
in $D'_m$. Since $a_0$ can be arbitrarily chosen, there are self-similarities
at all scales. Therefore, Riemann zeta function is a fractal.

All in all, by further choosing
\[f(s) = \zeta(a_{m}\: s \exp(i\:\theta_{0}) + s_{0}) \; \; \; \mbox{or}
\; \; \; f(s) = \zeta(i\: a_{m}\:  \bar{s} \exp(i \theta_{0}) + s_{0}) \]
where $\bar{s}$ is the conjugate of $s$, and
$a_m, \theta_0$ and $s_0$ are such that $f(s) \neq 0$ for $|s| \leq r$,
we have self-similarities between discs at different scales and orientations
(rescaled by $a_m$, and/or rotated by $\theta_0$ about the center of the disc
and/or, flipped or reflected by $i\: \bar{s}$ about any diagonal line at angle
$\theta_0$ crossing the center of the disc).

\bigskip

\noindent {\bf Remarks} \begin{enumerate}
\item Riemann zeta function is fractal in the sense that Mandelbrot set is
fractal (self-similarities between a region bounded by a closed loop $C$ and
other regions bounded by closed $C'_m$ of the same shape at smaller scales
and/or at different orientations).
The fractal property of zeta is not ``infinitely recursive'' as in Kock
Snowflake. Such infinite recursions in a function, ie.
$\dots \subseteq D'_n \subseteq D'_{n-1} \subseteq \dots \subseteq D'_1
\: \:$ or \\
$\dots = T_n = T_{n-1} = \dots = T_1$, \\
will render the function non-differentiable, whereas zeta is infinitely
differentiable. So, the manifold of zeta function is not of fractal dimension.

\item If we vary $a_m$ and $r$ such that $|a_m \: s| = r_{0}$, where
$r_0$ is an arbitrary fixed radius and $|s| \leq r < 1/4$, then every mapping
of $\zeta(a_m \:s+s_{0})$ for a disc of arbitrary radius $r_{0}$
containing no zeros of $zeta$ and centered anywhere in the complex plane (at
arbitrary $s_{0}$) are replicated by infinitely many discs at every radius
$\leq r$ centered along $Re(s') = 3/4$ within the right half of critical
strip as in {\bf Figure $2$}.

\begin{figure}[htbp]
\centerline{ \psfig{file=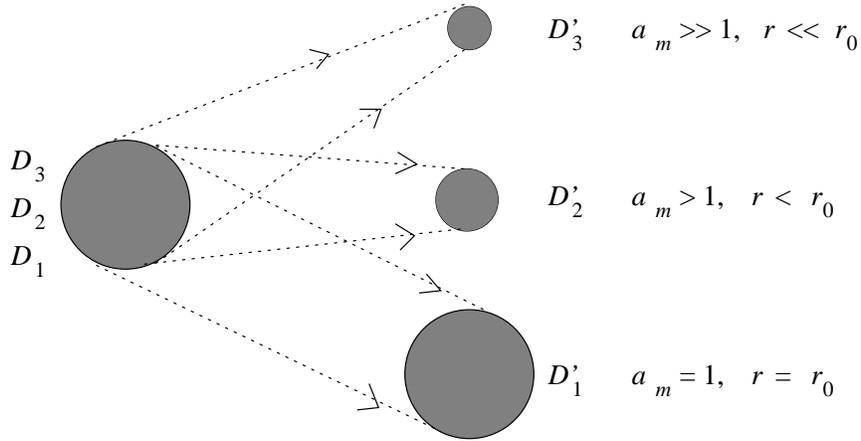}}
\caption{Map $\mu(m) : \; \; D_m \longmapsto D'_m$, where $a_m$ and $s$
are such that $|a_m \: s| = r_0$.}
\end{figure}

\item All Dirichlet L-functions are also fractal. This follows from the remark
following Voronin's theorem in Voronin's paper \cite{Voronin}.
\end{enumerate}

\bigskip

\begin{corollary} Riemann zeta function is a ``library'' of all possible
smooth continuous line drawings in a plane, so long as $f(s)$ is analytic and
$f(s) \neq 0$ for $|s| \leq r$. \end{corollary}

If we choose analytic function $f(s)$ to map a line in s to an amusing curve
(so long as $f(s)$ is analytic and $f(s) \neq 0$ for $|s| \leq r$) whose path
traces out the outline of a cartoon character, say Mickey Mouse, in the
complex plane (not much different from a loop), then by Voronin's theorem,
we can always find a finite $T$ in which $\zeta(s+3/4+i\:T)$
traces out the Mickey Mouse outline. Since we can substitute the Mickey Mouse
outline with any outline drawings, Riemann zeta function is a
{\em ``library''} containing all possible such drawings!

\begin{corollary} Riemann zeta function is a concrete ``representations''
of the giant book of theorems referred to by Paul Halmos. \end{corollary}

Now substitute the outline curve of $f(s)$ in {\bf Corollary 2} with a smooth
oscillating curve in complex plane. Let the oscillating curve have
periods or cycles varying in such a way that it carries a message encoded
in Morse codes (oscillating curve having the shape of the signal amplitude
vs time of a telegram transmission).
Let the message be the entire text of the Encyclopedia Brittanica.
By Voronin's theorem, we can always find a finite $T$ for which
$\zeta(s+3/4+i\:T)$ traces out out an arbitrarily good copy (errors within
$\epsilon$) of the oscillating curve carrying identical encoded messages.
So, similarly, the entire human knowledge are already encoded in zeta
function.

Since the oscillating curve can be arbitrarily rescaled without changing
the content of the encoded messages, there is, by {\bf Corollary 1}, an
infinite number of arbitrarily good copies of the messages encoded with the
same encoding scheme at different discs in Riemann zeta function.

Hence, Riemann zeta function is probably one of the most remarkable functions
because it is a concrete ``representation'' (in group theory sense) of
``the God's giant book of theorems'' that Paul Halmos spoke of --- all
possible theorems and texts are already encoded in some form in Riemann zeta
function, and repeated infinitely many times. Although a white noise function
and an infinite sequence of random digits are also concrete
``representations'', Riemann zeta function is not white noise or random but
well-defined.

Alternatively, from the point of view of information theory, even though
Riemann zeta function is well-defined, its mappings in the right half of the
critical strip are random enough to encode arbitrary large amount of
information --- the ``entropy'' of its mapping is infinite.

\noindent {\bf Example} \\
\indent This article is also encoded somewhere in Riemann zeta function as it
is being written!

\noindent {\em Acknowledgement} --- The author wishes to thank C. Davis,
R.P. Brent, and A.M. Odlyzko for helpful discussions.

\end{document}